\documentclass[aps,showpacs,amsmath,amssymb,12pt]{revtex4-1}

\usepackage{graphicx}
\usepackage{dcolumn}
\usepackage{bm}
\usepackage{subfigure}
\usepackage{longtable}
\makeatother

\begin{document}
\title{Geodesic Motion on Closed Spaces: Two Numerical Examples}

\author{Daniel M\"uller}
\affiliation{Universidade de Bras\'\i lia - Instituto de F\'\i sica, Cxp 04455, Asa Norte, 70919-900, Bras\'\i lia, DF, Brazil}
\email{ muller@fis.unb.br}

\begin{abstract}
The geodesic structure is very closely related to the trace of the Laplace operator, involved in the calculation of the expectation value of the energy momentum tensor in Universes with non trivial topology. The purpose of this work is to provide concrete numerical examples of geodesic flows. Two manifolds with genus $g=0$ are given. In one the chaotic regions, form sets of negligible or zero measure. In the second example the geodesic flow, shows the presence of measurable chaotic regions. The approach is ``experimental", numerical, and there is no attempt to an analytical calculation.
\end{abstract}
\maketitle
\section{Introduction}
\label{intro}
Anosov's famous result, states that the geodesic flow is chaotic in a compact manifold of constant negative curvature \cite{anosov}. Anosov flows are very chaotic being not only mixing, but even Bernoullian \cite{ber}. For instance, the Poincar\'e section shows the absence of KAM tori \cite{KAM}. For a review, see \cite{BV}. 

On the other hand integrability or not of a given mechanical system remains as an non answered question, albeit much progress had occurred recently \cite{livro}. Since Krylov's work \cite{Kr79} many researchers have transformed the mechanical problem of the motion of a particle in a given potential into a billiard problem \cite{cit}. This is achieved by writing the Jacobi metric associated to the given potential. In this approach, the motion is geodesic. The Jacobi manifold is specifically obtained to incorporate the effects of the fields.

In this context the Henon-Heiles Hamiltonian provides an example of an everywhere positive curvature space which is well known to be chaotic \cite{csp}. More generally, chaos is related to the parametric instability induced by variations of the scalar curvature along the geodesic \cite{ccp}. Also, the infinite hyperbolic space, is very well known to be integrable and an everywhere negative (constant) curvature space. There is not any rigorous relation between instability of the geodesic and the curvature of the manifold. 

In the cosmological context it is the geometry itself the more fundamental field.
The geodesic motion of particles, follow directly from the covariant divergence of the energy momentum tensor \cite{papapetrou}. A few years ago Cornish et al suggested that the chaotic motion of particles in a closed negatively curved manifold as a possible mechanism responsible for the homogenization of the Universe \cite{cornish}. 

Anyway for spatialy multiply connected manifolds, the point $x$ and the
point $\gamma x$ where $\gamma \in \Gamma$, $x \equiv \gamma x$ for all
the elements of the fundamental group $\Gamma$. That is, the point $x$
and $\gamma x$ are the same and identical point. This means that any function on this manifold must be periodic 
in some sense. Mathematically, the  functions defined on a closed 
manifold $\mathcal{M}$ are called automorfic, see for example 
\cite{balazsvoros}, \cite{BV}.

One way of obtaining this periodicity is by imposing summations over the spectrum of the Laplace operator of the particular manifold in question. It is
well known that summations over the spectrum are equivalent to
summation over the closed geodesics, also known as the method of
images, \cite{BV}, \cite{balazsvoros}. This result lies at the 
hart of the
Selberg formalism for the calculations of functional traces and is 
valid at least for manifolds, Lie groups included \cite{Camporesi}.

Also, the geodesic structure is very important in connection with quantum chaos, see for example \cite{ott}, \cite{BV}. The duality spectrum-geodesic was used to obtain the Casimir energy in closed Universes, by collaborators and myself for example \cite{daniel}.

In 1839 it was discovered by Jacobi itself, that the geodesic motion on an ellipsoid is integrable \cite{Jacobi}. Jacobi used a particular coordinate system \cite{Arnold}, now known as Jacobi elliptical coordinates, and obtained an independent, and involutive constant of the motion. In 1994, Knieper and Weiss \cite{KW} proved that there are many smooth Riemannian metrics on $S^2$ with chaotic geodesic flows. The authors consider arbitrary small deviations of the metric on the ellipsoid. Then, the Melnikov method is used to prove the existence of an homoclinic point. 

The purpose of this article is to provide a concrete example of two manifolds topologically equivalent to a sphere. The geodesics are investigated using the  technique of the Poincar\'e surface of section. Section \ref{s2} includes a very brief discussion of the Gauss-Bonnet theorem, for more details see \cite{dfn}.
In section \ref{s1} the geodesic motion of the given manifold, shows that the non integrable regions form a set with negligible or zero measure in phase space. In section \ref{s2} the geodesic flow on the manifold, shows that the chaotic regions are enlarged. We speculate that the presence of chaos is related to regions of positive and negative values of the scalar curvature for this particular example under consideration. 

Anyway, in Section \ref{s2} and in the conclusions it is stressed that a closed space with domains of negative and positive curvature is not a mandatory condition for chaotic geodesic motion.

\section{An everywhere positive curvature space}
\label{s1}
\begin{figure}[h]
\begin{center}
\includegraphics[scale=0.4]{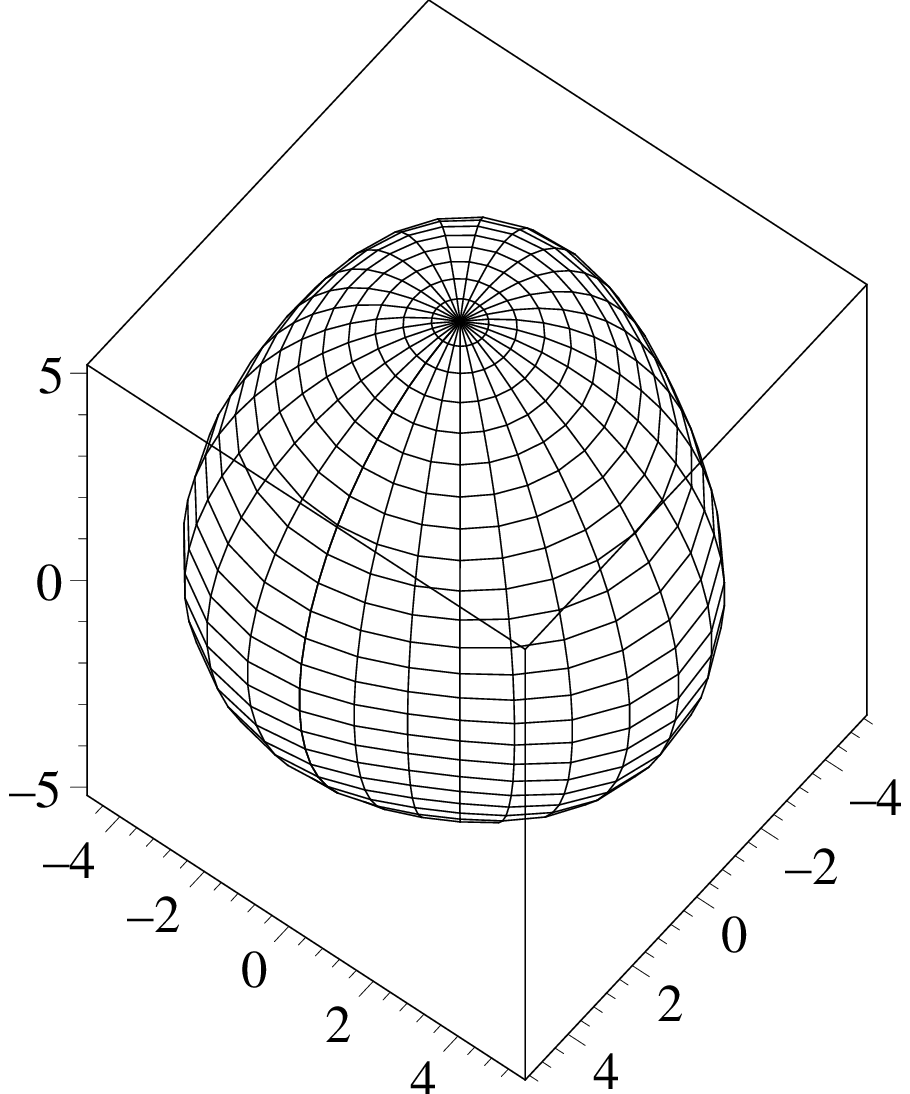}
\caption{The manifold, the parameter $a=1$ in (\ref{imersao}) is chosen.}
\label{f1}
\end{center}
\end{figure}

\begin{figure}[h]
\begin{center}
\includegraphics[width=7cm,height=6cm]{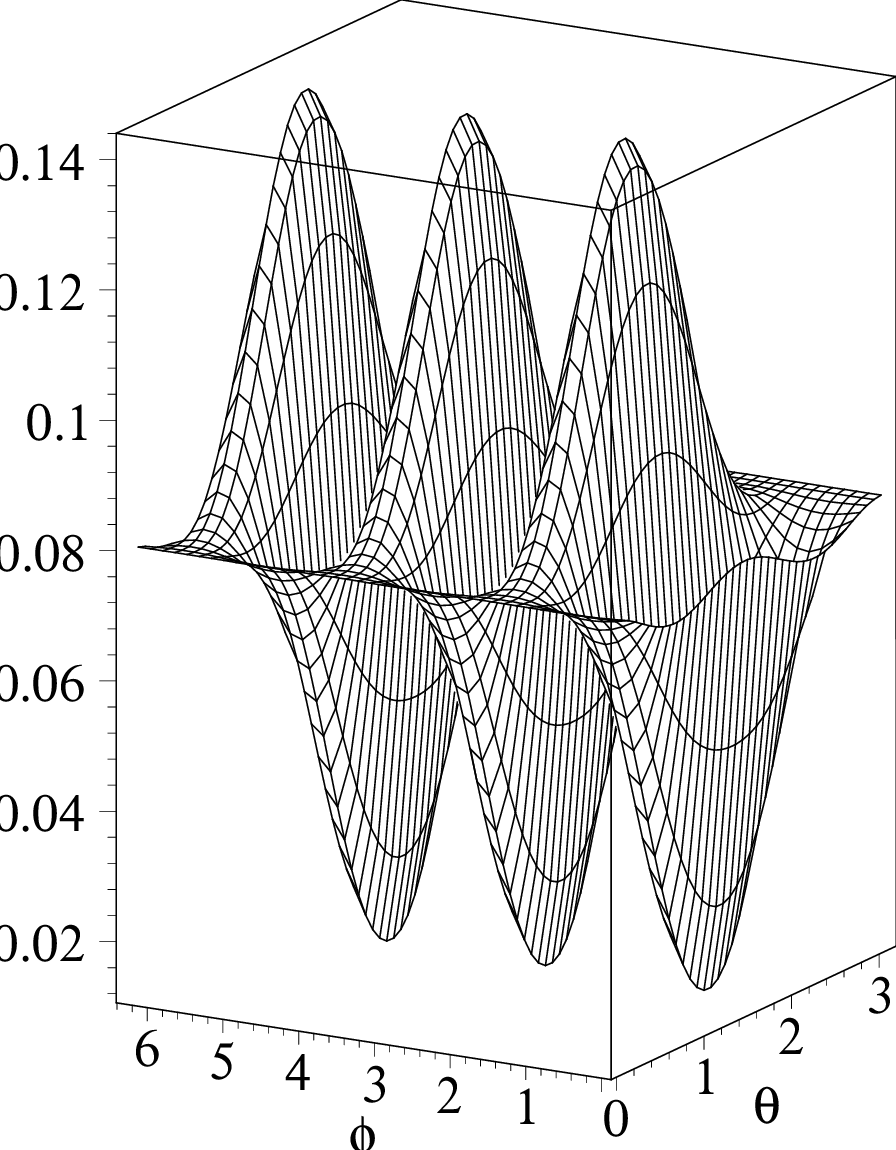}
\caption{Riemann scalar curvature for the metric in (\ref{metrica}) against $\theta,\,\phi$ for the manifold given in FIG.  \ref{f1}, with $a=1$.}
\label{Riemann}
\end{center}
\end{figure}

This manifold is obtained as the immersion of a closed surface in the Euclidean space $E^3$. As is well known, the spherical harmonics $Y^m_l(\theta,\phi)$ form a complete base for any function defined on $S^2$. In this work, the following class of surfaces
\begin{eqnarray}
&&r=5+aY^3_3(\theta,\phi)\nonumber\\
&&r=5-a\frac{\sqrt{35}\sin(3\phi)\sin(\theta)^3}{8\sqrt{\pi}}
\label{imersao}
\end{eqnarray}
is considered. When $a=0$, it corresponds to the usual $S^2$. For small values of $a$, it provides deformations of $S^2$. In this section the particular value $a=1$ is chosen, as shown in FIG.  \ref{f1}.
The line element is the usual one 
\begin{eqnarray*}
&&ds^2=dr^2+r^2(d\theta^2+\sin(\theta)^2d\phi^2)\\
&&ds^2=\left( \frac{\partial}{\partial\theta}rd\theta+\frac{\partial}{\partial\phi}rd\phi\right)^2+\left(5-a\frac{\sqrt{35}\sin(3\phi)\sin(\theta)^3}{8\sqrt{\pi}}\right)^2(d\theta^2 +\sin(\theta)^2d\phi^2),
\end{eqnarray*} 
and the induced metric on the surface 
\begin{eqnarray}
&&g=\nonumber\\
&&\hspace{-2.5cm}
\left[ \begin {array}{cc} {\frac {315}{64}}\,{\frac {{a}^{2} \left( \sin \left( \theta \right)  \right) ^{4} \left( \sin \left( 3\,\phi
 \right)  \right) ^{2} \left( \cos \left( \theta \right)  \right) ^{2}
}{\pi }}+ \left( {\frac {40\sqrt{\pi}-a\sqrt {35} \left( \sin \left( \theta
 \right)  \right) ^{3}\sin \left( 3\,\phi \right) }{8\sqrt {\pi }}}
 \right) ^{2}&{\frac {315}{64}}\,{\frac {{a}^{2} \left( \sin \left( 
\theta \right)  \right) ^{5}\sin \left( 3\,\phi \right) \cos \left( 
\theta \right) \cos \left( 3\,\phi \right) }{\pi }}
\\\noalign{\medskip}{\frac {315}{64}}\,{\frac {{a}^{2} \left( \sin
 \left( \theta \right)  \right) ^{5}\sin \left( 3\,\phi \right) \cos
 \left( \theta \right) \cos \left( 3\,\phi \right) }{\pi }}&{\frac {
315}{64}}\,{\frac {{a}^{2} \left( \sin \left( \theta \right)  \right) 
^{6} \left( \cos \left( 3\,\phi \right)  \right) ^{2}}{\pi }}+ \left( 
{\frac {(40\sqrt{\pi}-a\sqrt {35} \left( \sin \left( \theta \right)  \right) 
^{3}\sin \left( 3\,\phi \right))\sin(\theta) }{\sqrt {8\pi }}} \right) ^{2} 
\end {array} \right]\nonumber\\
&&
\label{metrica}
\end{eqnarray}
The Riemann scalar curvature for the metric (\ref{metrica}), with $a=1$ is shown in FIG.  \ref{Riemann}. 

Given a geodesic Lagrange function
\[
L=\frac{1}{2}g_{ab}\dot{x}^a\dot{x}^b,
\]
where $\dot{x}^a=\frac{d}{d\lambda}x^a$ and $\lambda$ is a parameter along the geodesic, the Hamiltonian follows 
\[
H=\frac{1}{2}g^{ab}p_ap_b,
\]
where $p_a$ are the momenta conjugate do the velocity $\dot{x}^a$. For the case of interest we have
\begin{eqnarray}
H=\left\{\left(\frac{315}{64}\,\frac{a^{2}\left(\sin\left(\theta\right)\right)^{6}\left(\cos\left(3\,\phi\right)\right)^{2}}{\pi}+r^{2}\left(\sin\left(\theta\right)\right)^{2}\right)p_\theta^{2}\right.\nonumber\\
\left.-\frac{315}{32}\,\frac{a^{2}\left(\sin\left(\theta\right)\right)^{5}\sin\left(3\,\phi\right)\cos\left(\theta\right)\cos\left(3\,\phi\right)p_\theta\,p_\phi}{\pi}\right.\nonumber\\
\left.+\left(\frac{315}{64}\,\frac{a^{2}\left(\sin\left(\theta\right)\right)^{4}\left(\sin\left(3\,\phi\right)\right)^{2}\left(\cos\left(\theta\right)\right)^{2}}{\pi}+r^{2}\right)p_\phi^{2}\right\}\nonumber\\
\left\{\left(\frac{315}{64}\,\frac{a^{2}\left(\sin\left(\theta\right)\right)^{4}\left(\sin\left(3\,\phi\right)\right)^{2}\left(\cos\left(\theta\right)\right)^{2}}{\pi}+r^{2}\right)\right.\nonumber\\
\left.\left(\frac{315}{64}\,\frac{a^{2}\left(\sin\left(\theta\right)\right)^{6}\left(\cos\left(3\,\phi\right)\right)^{2}}{\pi}+r^{2}\left(\sin\left(\theta\right)\right)^{2}\right)\right.\nonumber\\
\left.-\frac{99225}{4096}\,\frac{a^{4}\left(\sin\left(\theta\right)\right)^{10}\left(\sin\left(3\,\phi\right)\right)^{2}\left(\cos\left(\theta\right)\right)^{2}\left(\cos\left(3\,\phi\right)\right)^{2}}{\pi^{2}}\right\}^{-1},
\label{hamiltoniano}
\end{eqnarray}
where $a=1$, and $r$ is given by the equation (\ref{imersao}). As we are considering geodesics, the value of the energy $H=E$ can be easily rescaled and absorbed into a new time scale. For instance the energy can be set to one $E=1$, then the parameter $\lambda$ is equivalent to the geodesic distance $s$, $\lambda\equiv s$. Since the value of the energy can be trivially rescaled it was chosen at will throughout this work. Of course the Poincar\'e sections refer to fixed, constant energy surfaces. 

We quote here the geodesic deviation equations 
\[
u^a\nabla_an^b=R^b_{klm}u^ku^ln^m,
\]
where $R^b_{klm}$ is the Riemann tensor for the metric in (\ref{metrica}).
Which in Fermi coordinates $E_1=u$, $E_1$ is in the direction of the velocity vector $u$, and 
$E_2=xn$, where $x$ is the orthogonal separation between neighbouring geodesics
\begin{equation}
\frac{d^2x}{d\lambda^2}=-Kx,
\label{dgeodesico}
\end{equation}
where $K$ is Gauss's curvature of the surface. The Gaussian curvature $K$ of a surface, is given by the ratio of the determinant of the second fundamental form, by the first fundamental form. The relation between the Gaussian curvature and the Riemann scalar is very simple 
\[
R=2K.
\]
If the curvature is positive $K>0$, we can see that the solutions of (\ref{dgeodesico}) do not diverge.

\begin{figure}[h]
\begin{center}
\includegraphics[scale=0.6,angle=-90.0]{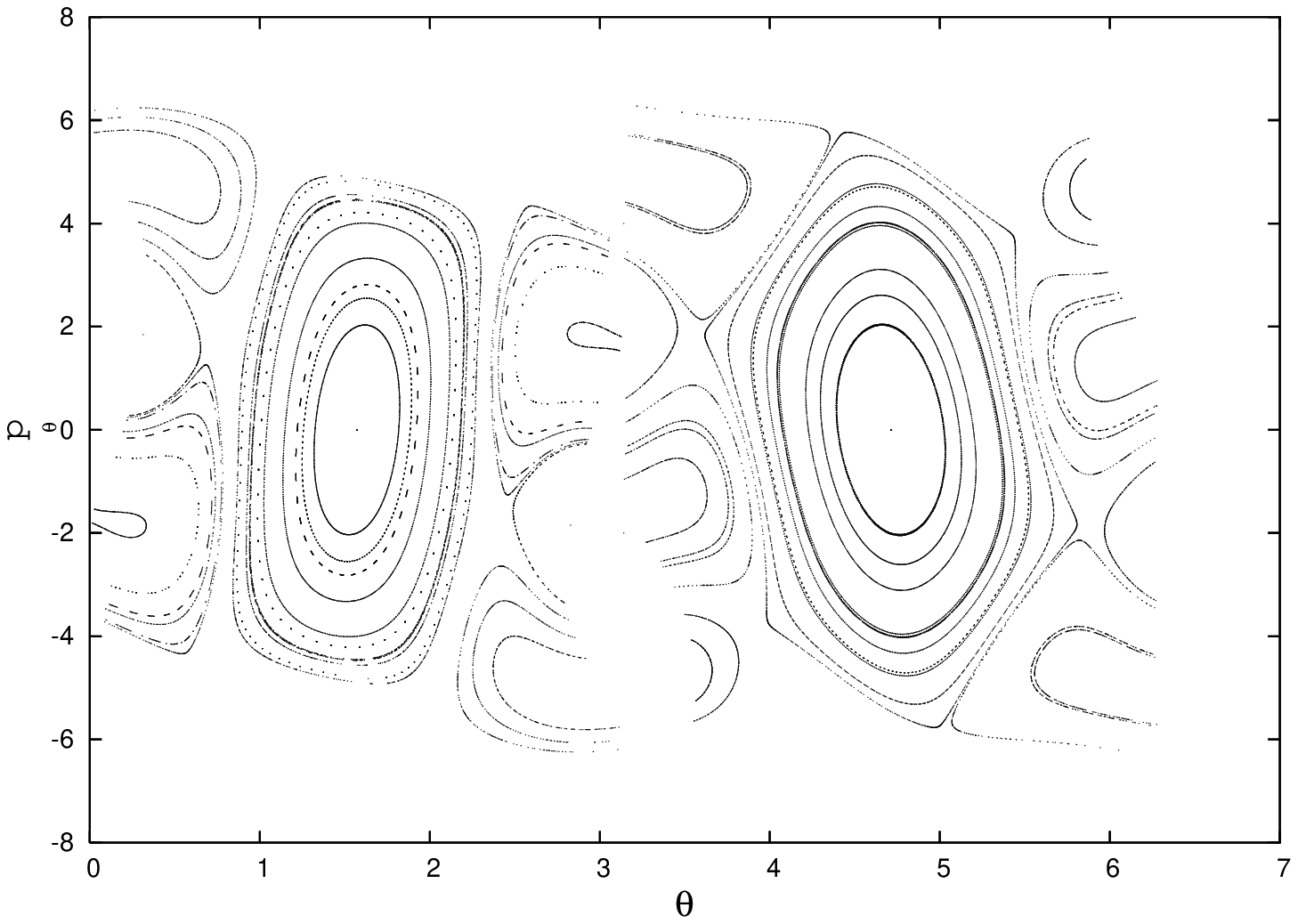}
\caption{Poincar\'e section for the geodesic flow given by (\ref{hamiltoniano}) with $a=1$. The intersection surface is set to $\phi=5.0$ and the constant Hamiltonian shows a cumulative error in one part in $10^{11}$. The non integrable regions, if any, are small.}
\label{int}
\end{center}
\end{figure}

The Poincar\'e section for the Hamiltonian (\ref{hamiltoniano}) with $a=1$, is shown in Fig \ref{int}. The intersection surface is set to $\phi=5.0$ and the constant Hamiltonian, with $H=1.98765$ shows a cumulative error in one part in $\sim 10^{11}$. If the system is integrable the non integrable regions are identically of zero measure. Visually, the non integrable regions, if any, are of ``small" measure in FIG.  \ref{int}.

\newpage
\section{A manifold with domains of negative curvature}
\label{s2}
The immersed surface is very similar as the one in the last section, given by (\ref{imersao}); the difference is that now $a=2$. The metric is the same (\ref{metrica}), with $a=2$, and the Hamiltonian also, (\ref{hamiltoniano}), with 
$a=2$.

\begin{figure}[h]
\begin{center}
\includegraphics[scale=0.4]{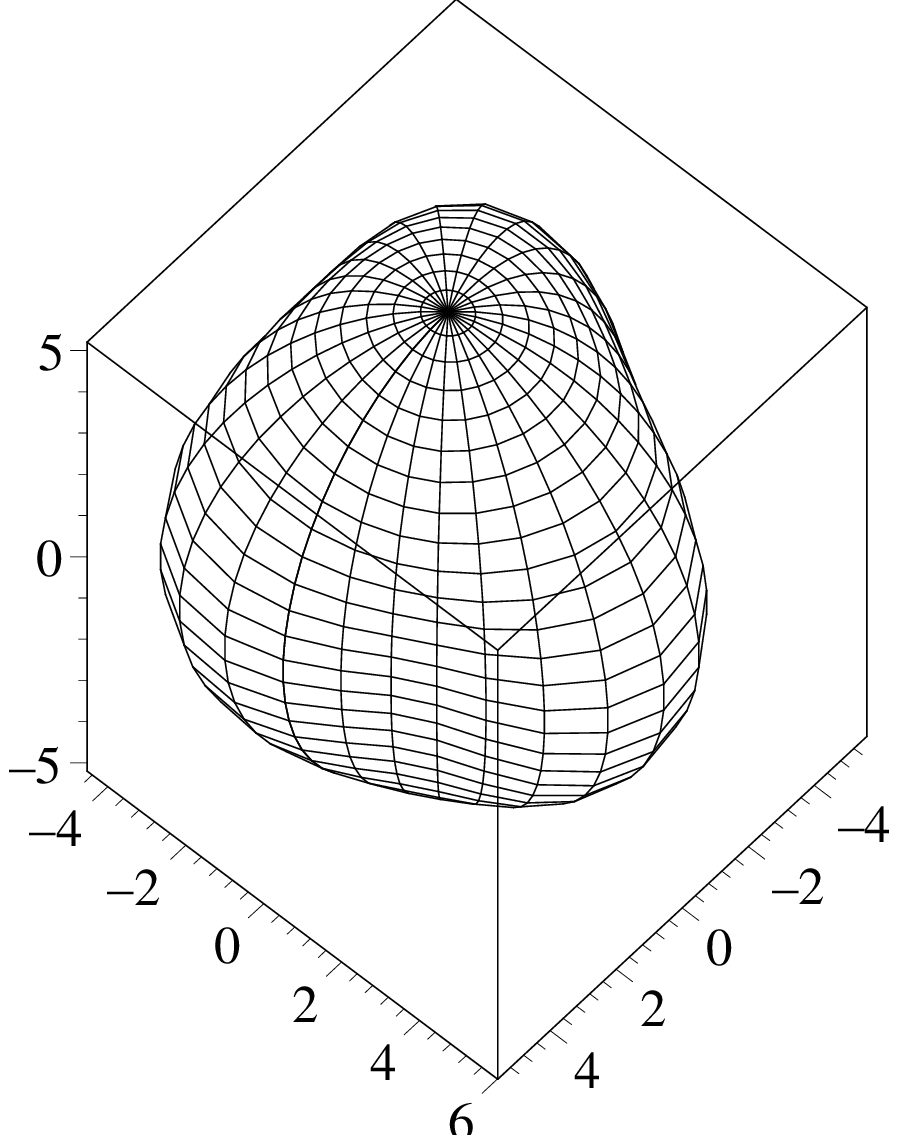}
\caption{The manifold. The parameter $a=2$ in (\ref{imersao}) is chosen.}
\label{f2}
\end{center}
\end{figure}

\begin{figure}[h]
\begin{center}
\includegraphics[width=7cm,height=6cm]{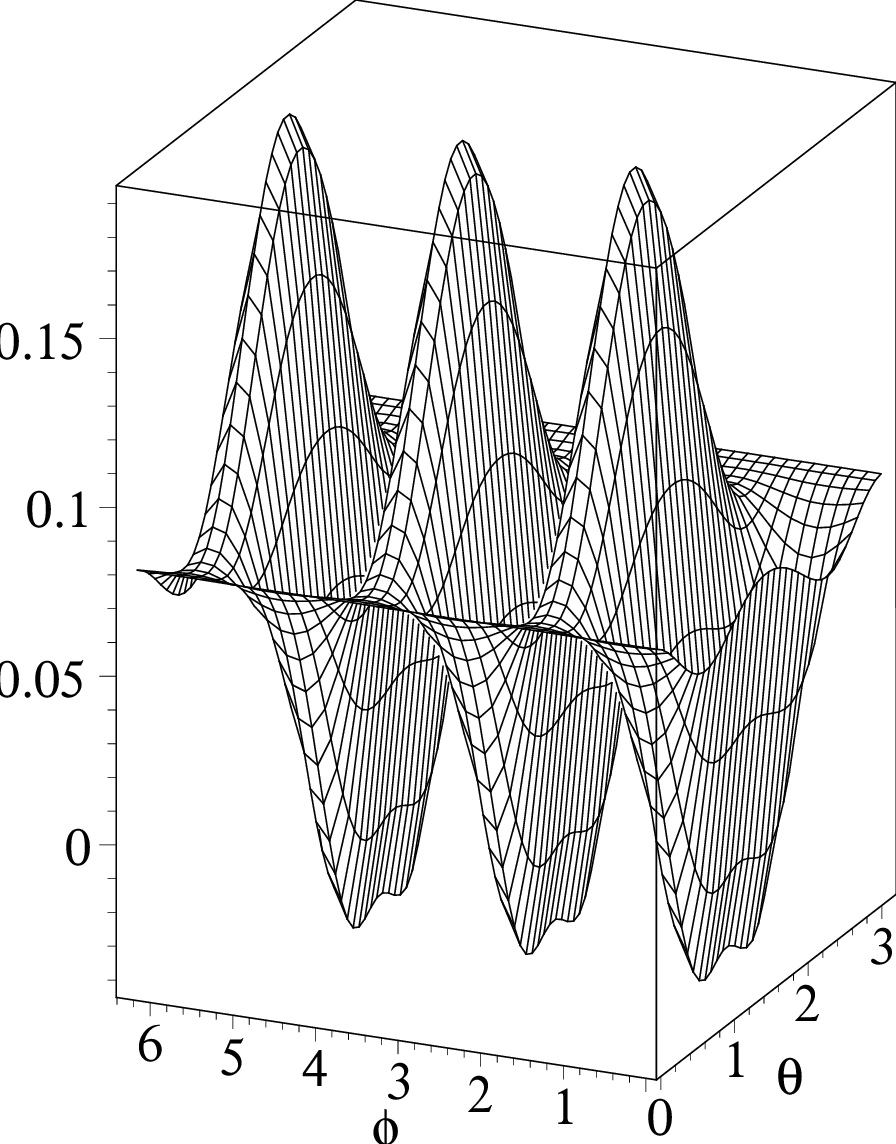}
\caption{Riemann scalar curvature for the metric given in (\ref{metrica}), against $\theta,\,\phi$ for the manifold given in FIG.  \ref{f2}, with $a=2$.}
\label{Riemann2}
\end{center}
\end{figure}

The Gauss-Bonnet theorem relates the integrated Gaussian curvature to the Euler number of the surface $\chi=2-2g$, for more details see \cite{dfn}. Of course, both manifolds in this work, have $g=0$, as can be seen in FIG.  \ref{f1} and FIG.  \ref{f2}. Anyway this second manifold has some domains with negative curvature as can be seen by Riemann scalar curvature for the metric given in (\ref{metrica}) plotted in FIG.  \ref{Riemann2}. 

According to (\ref{dgeodesico}) in these domains, since $K<0$, the geodesics diverge exponentially. 

\begin{figure}[h]
\begin{center}
\includegraphics[scale=0.6,angle=-90.0]{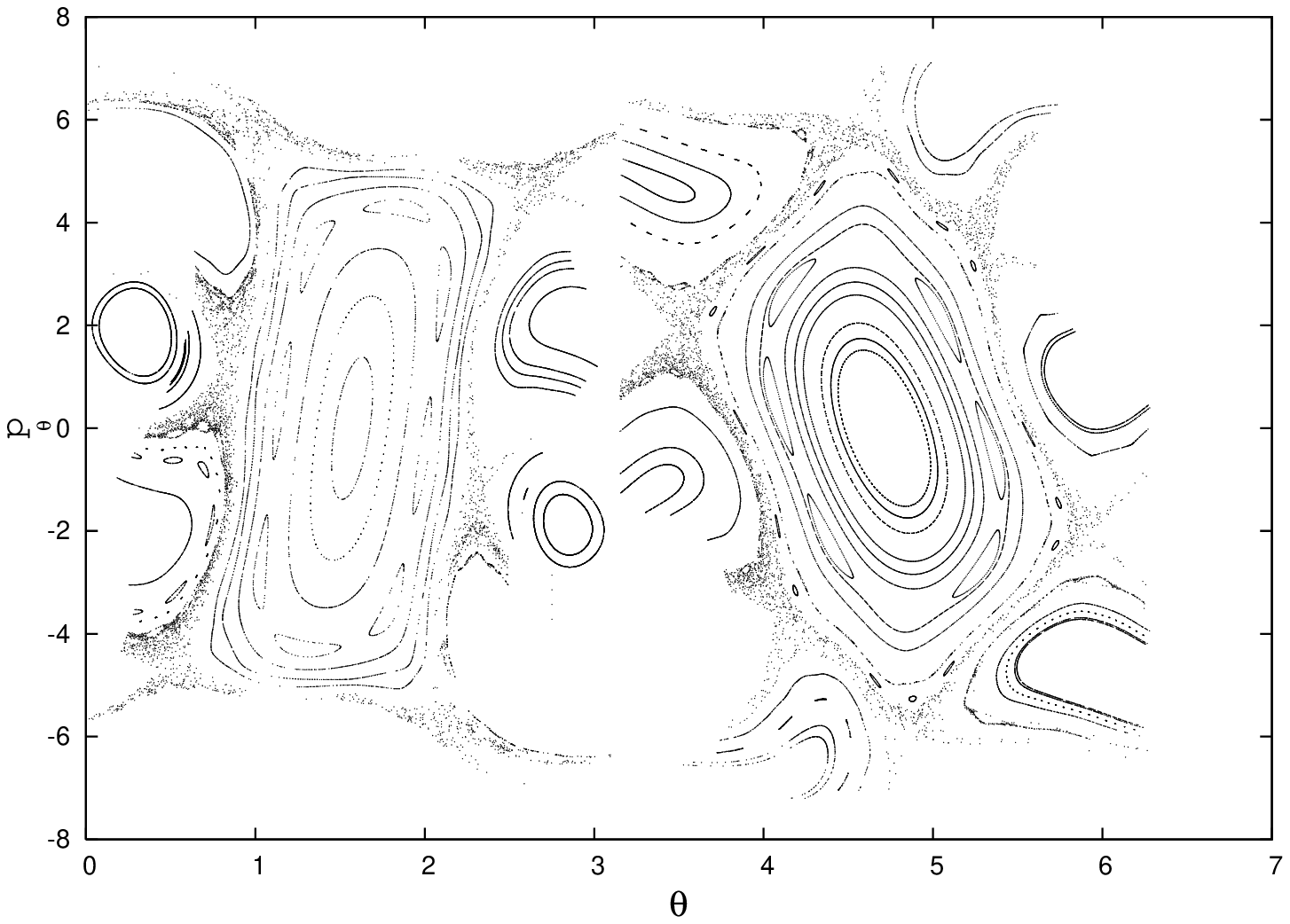}
\caption{Poincar\'e section for the geodesic flow given by (\ref{hamiltoniano}), with $a=2$. The intersection surface is set to $\phi=5.0$ and the constant Hamiltonian, shows a cumulative error in one part in $\sim 10^{11}$.}
\label{caos}
\end{center}
\end{figure}

The Poincar\'e section for the Hamiltonian (\ref{hamiltoniano}) with $a=2$, is shown in FIG.  \ref{caos}. The intersection surface is set to $\phi=5.0$ and the constant Hamiltonian, with $H=1.98765$ shows a cumulative error in one part in $\sim 10^{11}$. The system is not integrable as the chaotic regions can be seen in FIG.  \ref{caos}. 

\begin{figure}[h]
\begin{center}
\includegraphics[scale=0.6]{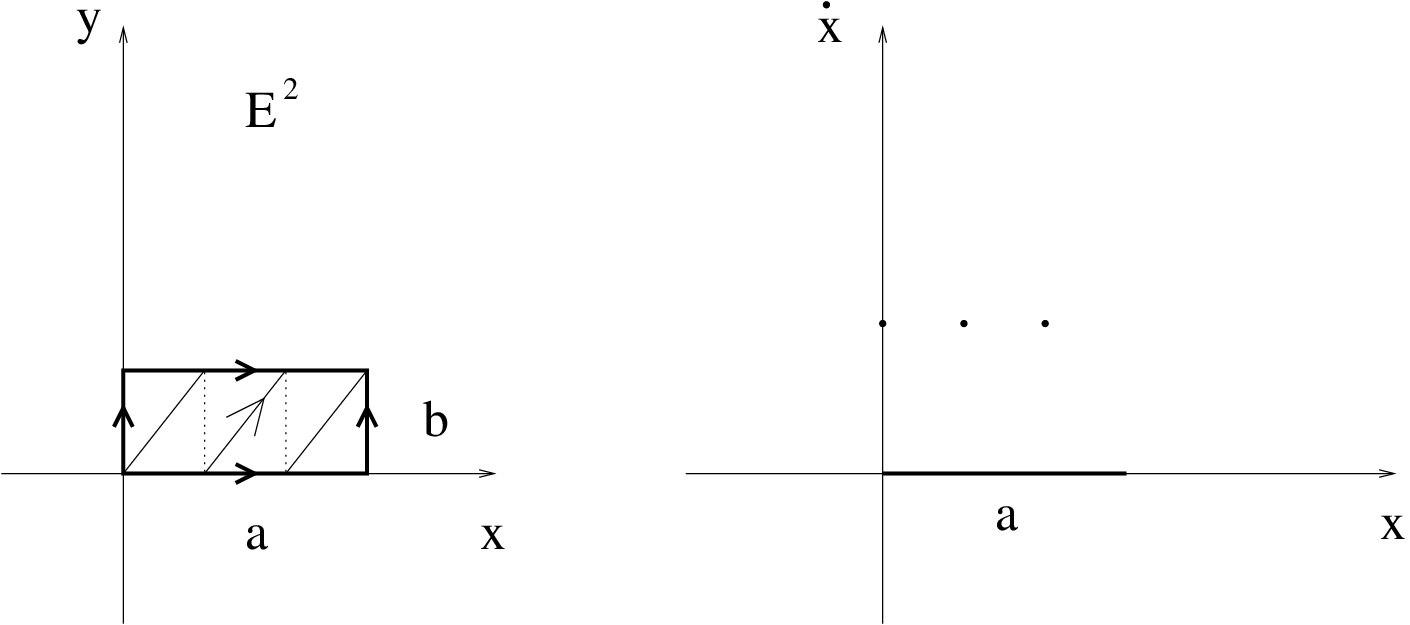}
\caption{On the left it is shown the genus $g=1$ torus with its fundamental region in bold, and the covering space, the Euclidean space $E^2$. It is also shown, the generators of the fundamental group, the discrete translations $a$ and $b$ and a periodic geodesic. It is shown the Poincar\'e section for the periodic geodesic drawn on the right. The intersection surface is set to $y=0$.}
\label{toro}
\end{center}
\end{figure}

For the case of the torus with genus $g=1$, there are regions of negative Gaussian curvature $K<0$ and positive Gaussian curvature $K>0$. The geodesic motion on the torus is integrable.
Let us clarify what the above means. The fundamental group of the torus is 
$\pi_1=\mathbb{Z}\otimes \mathbb{Z}$, which corresponds to discrete translations $a$, $b$, FIG. \ref{toro}. In FIG.  \ref{toro} it is also shown a periodic geodesic on the torus, with period $3:1$. On the right side of FIG.  \ref{toro} it is shown the surface of section with the plane $y=0$, corresponding to the periodic orbit. There are only 3 points. This occurs because all the elements of the fundamental group, commute with the linear momentum $\vec{p}=(\dot{x},\dot{y})$. For a non periodic orbit, i.e. a periodic orbit with a period $T\rightarrow \infty$, the surface of section for the single orbit would be a ``continuous" line. This line corresponds to a surface in phase space, that is, a zero measure set in the constant energy $3-$D phase space. This motion is not ergodic in the constant energy allowable phase space.

Geometrically a compact space of constant negative curvature is a torus with genus $g>1$. In this case, the Poisson bracket of the infinitesimal generators of the fundamental group with the momenta are not zero $\{\xi_i,\vec{p}\}\neq 0$, i.e. they do not commute. This is what originates chaos in closed, smooth, constant negatively curved manifolds, for more details see \cite{BV}.

\section{Conclusions}
Anosov's famous result, states that the geodesic flow is chaotic in a compact manifold of constant negative curvature \cite{anosov}. Anosov flows are very chaotic being not only mixing, but even Bernoullian \cite{ber}. For instance, the Poincar\'e section shows the absence of KAM tori \cite{KAM}. For a review, see \cite{BV}. 

Integrability or not of a given mechanical system remains as an non answered question. Since Krylov's work \cite{Kr79} many researchers have transformed the mechanical problem of the motion of a particle in a given potential into a billiard problem \cite{cit}. This is achieved by writing the Jacobi metric associated to the given potential. In this approach, the motion is geodesic. The Jacobi manifold is specifically obtained to incorporate the effects of the fields.

There has been an attempt to establish a local criteria for chaos. In particular A. Saa \cite{cit} shows an example of chaos occurring in a strictly positive curvature space, $K>0$. J. Szcz\c esny and T. Dobrowolski in \cite{cit} show an example of an integrable system with $K<0$, namely the classical scattering Kepler problem. In the above mentioned examples the manifolds are not closed. 
This result is in contrast to the geodesic motion on $S^2$, $K=1$ which is integrable, and on the genus $g=2$ torus, $K=-1$  which is strongly chaotic \cite{anosov}. 

As it is well known, there is not any trivial relation between integrability or not of the geodesic flow to the sign of Gaussian curvature of the manifold.

In FIG. \ref{caos}, there is some geodesic chaos. The manifold which contains the geodesics is given in FIG.  \ref{f2} and it has regions with $K<0$ and $K>0$. According to equation (\ref{dgeodesico}), positive curvature approaches neighboring geodesics, and negative curvature diverge neighboring geodesics. 

More generally, chaos is related to the parametric instability induced by variations of the scalar curvature along the geodesic \cite{ccp}. In FIG.  \ref{int} the chaotic regions of the geodesic flow are small, and the curvature $K>0$ is positive everywhere. We have specifically checked smaller values of the parameter $a$ \eqref{metrica} so that the curvature is everywhere positive but very small for particular values of $\theta$ and $\phi$, for example $a=1.2$. For this value of $a=1.2$, still the chaotic regions are all very small. 
Apparently, only when there are negative values of $K$ the difference between their maximum and minimum values is enough to cause the desired instability. 

It is also well known that parametric instability can stabilize an unstable fixed point, see for instance the reversed pendulum \cite{Arnold} . The Jacobi manifold in this case exhibits mixed sign curvature. So mixed signed curvature can either result in ordered motion, otherwise it can result in chaotic motion. 

For instance, the fundamental region of the genus $g=1$ torus in FIG.  \ref{toro} is a rectangle with opposite sides identified. The geodesic flow in the torus is integrable. Of course the integrability of the torus is due to the involutive integrals of the motion, and yet provides an example in which mixed sign curvature does not introduce instabilities. 

\acknowledgements
D. M. wishes to thank the Brazilian projects: {\it Nova F\'\i sica no Espa\c co} and INCT-A. The author wishes to thank an anonymous referee for many improvements and corrections.


\begin{thebibliography}{99}
\bibitem{anosov}D. V. Anosov, Proc. Steklov Math. Inst. {\bf 90}, 1, (1967).
\bibitem{ber}D. Ornstein and B. Weiss, Israel J. Math. {\bf 14}, 184, (1973).
\bibitem{KAM}A. N. Kolmogorov, Dokl. Akad. Nauk. SSSR, 98, 527 (1954); V. I. Arnold, Soviet Math. Dokl., 2, 501 (1961); J. Moser, Nachr. Akad. Wiss.  
G\"otingen, Math. Phys. K1, p.1 (1962).
\bibitem{BV} N. L Balazs and A. Voros, Phys. Rept.{\bf 143}, 109 (1986);
V. I. Arnold and A. Avez, {\it Probl\`emes Ergodiques de la M\'ecanique Classique}, Gauthier-Villars, Paris, (1967).
\bibitem{livro}M. Pettini, {\it Geometry and Topology in Hamiltonian Dynamics and Statistical Mechanics}, IAM Series n.33, Springer, NY (2007).
\bibitem{Kr79} N. S. Krylov, {\it Works on the Foundations of Statistical Physics}, Princeton Univ. Press, Princeton, NJ, (1979). 
\bibitem{cit} A. Saa, Ann. Phys. (NY), {\bf 314}, 508 (2004); 
M. Robnik, J. Phys. {\bf A16}, 3971 (1983); J. Zscz\c esny and T. Dobrowoslki, Ann. Phys. (NY) {\bf 277}, 161 (1999);L. Casetti, M. Pettini and E.G.D. Cohen, Phys. Rept. {\bf 337}, 237 (2000).
\bibitem{csp} M. Cerruti-Sola and M. Pettini, Phys. Rev. {\bf E 53}, 179 (1996); M. Pettini and R. Valdettaro, Chaos {\bf 5}, 646 (1995).
\bibitem{ccp} L. Casetti, C. Clementi and M. Pettini, Phys. Rev. {\bf E54}, 5969 (1996). 
\bibitem{papapetrou} A. Papapetrou, Proc. Roy. Soc. Lond., {\bf A209}, 248, (1951).
\bibitem{cornish} Cornish N. J., Spergel D., Starkman G., Phys. Rev.
Lett. {\bf 77}, 215 (1996).
\bibitem{balazsvoros} D. A. Hejhal, Duke Mathematical Journal,
  \textbf{43}, 441 (1976); A. B. Venkov, Russian Math. Surveys
  \textbf{34}, 79 (1979); R. Banach and J. S. Dowker, J. Phys. A
  \textbf{12}, 2545 (1979); R. Banach and J. S. Dowker, J. Phys. A 
\textbf{12}, 2527 (1979).
\bibitem{Camporesi} R. Camporesi, Phys. Rept. \textbf{196}, 1 (1990); 
J. S. Dowker Annals Phys. 62, 361 (1971).
\bibitem{ott} E. Ott, {\it Chaos in Dynamical Systems}, 2nd edition, Cambridge (2002).
\bibitem{daniel} D. Muller, H. V. Fagundes, R. Opher, Phys. Rev. {\bf D63} 123508 (2001); D. Muller, H. V. Fagundes, R. Opher, Phys. Rev. {\bf D66} 083507 (2002); D. Muller, H. V. Fagundes, Int. J. Mod. Phys.  {\bf A17}, 4385 (2002); M. P. Lima and D. Muller Class. Quant. Grav. {\bf 24}, 897 (2007).
\bibitem{Jacobi} C. G. J. Jacobi, Crelles J. {\bf 19}, 309, (1839).
\bibitem{Arnold} V. I. Arnold, {\it Mathematical Methods of Classical Mechanics}, Springer-Verlag, New York, (1978).
\bibitem{KW} G. Knieper and H. Weiss, J. Diff. Geom. {\bf 39}, 229, (1994).
\bibitem{dfn} B. Doubrovine, S. Novikov, A. Fomenko {\it G\'eom\'etrie Contempoirane - Methodes et applications 1e Partie} Mir, Moscou (1982); B. Doubrovine, S. Novikov, A. Fomenko {\it G\'eom\'etrie Contempoirane - Methodes et applications 2e Partie} Mir, Moscou (1982) 
\end{thebibliography}
\end{document}